\documentclass[reprint,
 prl,
 superscriptaddress,
 amsmath,
 amssymb,
 aps
]{revtex4-2}
\usepackage{lmodern}
\usepackage{graphicx}
\usepackage{dcolumn}
\usepackage{bm}
\usepackage{siunitx}
\usepackage{gensymb}
\usepackage{newtxtext}
\usepackage{newtxmath}
\usepackage{xcolor}
\usepackage[colorlinks=true]{hyperref} 

\AtBeginDocument{\def\selectlanguage#1{}}

\begin{document}

\title{Engineering and Probing a One-Dimensional Dipolar Spin Ensemble in Diamond}

\author{Lingjie Chen}
\thanks{These authors contributed equally to this work.}
\affiliation{Department of Physics, University of California, Santa Barbara, CA 93106, USA}

\author{Shreyas Parthasarathy}
\thanks{These authors contributed equally to this work.}
\affiliation{Department of Physics, University of California, Santa Barbara, CA 93106, USA}

\author{Simon A. Meynell}
\thanks{These authors contributed equally to this work.}
\affiliation{Department of Physics, University of California, Santa Barbara, CA 93106, USA}

\author{Lillian B. Hughes Wyatt}
\affiliation{Materials Department, University of California, Santa Barbara, CA 93106, USA}
\affiliation{Division of Engineering and Applied Science, California Institute of Technology, Pasadena, CA 91125, USA}

\author{Eveline Postelnicu}
\affiliation{Department of Materials Science and Engineering, Stanford University, Stanford, CA 94305, USA}

\author{Haopu Yang}
\affiliation{Department of Physics, University of California, Santa Barbara, CA 93106, USA}
\affiliation{Department of Physics, Harvard University, Cambridge, MA 02138, USA}

\author{Zilin Wang}
\affiliation{Department of Physics, Harvard University, Cambridge, MA 02138, USA}

\author{Weijie Wu}
\affiliation{Department of Physics, Harvard University, Cambridge, MA 02138, USA}

\author{Winston V. Peloso}
\affiliation{Department of Physics, University of California, Santa Barbara, CA 93106, USA}

\author{Casey K. Kim}
\affiliation{Materials Department, University of California, Santa Barbara, CA 93106, USA}

\author{Chris R. Laumann}
\affiliation{Department of Physics, Boston University, Boston, MA 02215, USA}
\affiliation{Max-Planck-Institut f\"ur Physik Komplexer Systeme, 01187 Dresden, Germany}

\author{Kunal Mukherjee}
\affiliation{Department of Materials Science and Engineering, Stanford University, Stanford, CA 94305, USA}

\author{Norman Y. Yao}
\affiliation{Department of Physics, Harvard University, Cambridge, MA 02138, USA}

\author{Ania C. Bleszynski Jayich}
\email[Corresponding author: ]{ania@physics.ucsb.edu}
\affiliation{Department of Physics, University of California, Santa Barbara, CA 93106, USA}

\date{\today}

\begin{abstract}
Dimensionality plays a central role in determining the collective behavior of interacting quantum systems.
Engineering strongly interacting ensembles of solid-state spin defects in reduced dimensions remains a significant challenge at the interface between the applied and fundamental sciences.
Here, we create and characterize a positionally disordered, quasi-one-dimensional spin chain in diamond, consisting of optically dark substitutional nitrogen defects (P1 centers) and optically addressable probe nitrogen-vacancy (NV) centers.
Our approach exploits the preferential incorporation of nitrogen along step bunches formed during chemical vapor deposition to achieve both lateral and vertical confinement.
Combining spatially resolved materials characterization with nanoscale quantum sensing, we establish the one-dimensional character of the optically dark, unpolarized P1 spin ensemble.
We then use correlation spectroscopy to probe local spin autocorrelations and investigate infinite-temperature dipolar spin transport.
Our results establish a materials-based route for engineering low-dimensional quantum spin systems.
\end{abstract}
\maketitle

Dimensionality plays a critical role in the physics of many-body quantum systems.
Reduced dimensionality can lead to enhanced quantum fluctuations, Luttinger liquid behavior, and unconventional phase transitions~\cite{mermin_absence_1966,hohenberg_theory_1977,sachdev_quantum_2023,giamarchi_quantum_2003}. 
One-dimensional quantum magnets have long provided a paradigmatic playground for exploring this wealth of phenomena in both theory and experiment~\cite{hayashi_nmr_1975,affleck_quantum_1989,choi_colloquium_2019,li_imaging_2024,joshi_observing_2022,wei_quantum_2022, peng_exploiting_2023}. 
On the experimental front, 1D quantum spin chains can naturally emerge in magnetic insulators~\cite{tanaka_electron_1985}, where localized moments are coupled through exchange interactions.
In such systems, recent works have centered on observations of anomalous transport, including, for example, super-diffusive spin transport in KCuF$_3$~\cite{scheie_detection_2021} and ballistic heat transport in SrCuO$_2$~\cite{hlubek_ballistic_2010}. 
Although magnetic materials provide a powerful setting for probing low-dimensional quantum dynamics, the preparation of tailored initial states is difficult, their microscopic Hamiltonians are largely determined by the host material, and experimental access is typically restricted to bulk observables.
Ensembles of color centers offer an intriguing alternative, combining many-body spin dynamics with state preparation, the ability to modify Hamiltonians via a well-developed pulse engineering toolbox, and refined spatial resolution via optical readout.
While recent progress has demonstrated the realization of strongly-interacting, two-dimensional color centers~\cite{davis_probing_2023,gong_coherent_2023,hughes_strongly_2025}, extending such systems to the one-dimensional regime has remained an outstanding challenge.

\begin{figure}
    \includegraphics{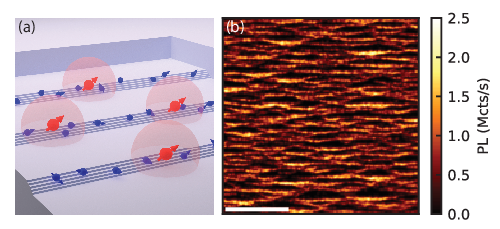}
    \caption{
    (a) Schematic showing nitrogen spins (blue) incorporated on step bunches on a (100)-oriented diamond surface during CVD growth. 
    Optically addressable NV centers (red) are used to probe the dynamics and dimensionality of the nitrogen spins through their mutual magnetic dipolar coupling, indicated by the translucent red spheres. 
    (b) Confocal photoluminescence (PL) image of the diamond under \SI{532}{\nano\meter} excitation, taken on the three-layer sample L043.
    Bright striations indicate spatially inhomogeneous NV density.
    Scale bar is \SI{10}{\micro\meter}.
    }
    \label{fig:Introduction}
\end{figure}

In this Letter, we take a critical step towards this goal and realize a one-dimensional interacting nitrogen (P1) spin ensemble in diamond.
Our main results are threefold.
First, we confirm, via spatially localized characterization of both defect density and surface topography, that surface morphology strongly controls the incorporation of nitrogen during diamond chemical vapor deposition (CVD)~\cite{de_theije_effects_2000, achard_coupled_2007, chayahara_effect_2004}.
In particular, the presence of step-edge features naturally leads to the generation of quasi-one-dimensional spin chains composed of optically active nitrogen-vacancy (NV) centers and dark P1 centers.
Second, we complement our materials-based characterization by analyzing NV center decoherence under driven P1 dynamics, allowing us to probe the nanoscale geometry of the P1 ensemble in diamond and confirming its one-dimensional nature. 
Finally, by measuring local autocorrelation functions of the P1 bath, we investigate the dipolar-interaction-driven dynamics of our disordered 1D spin ensemble, observing dynamics consistent with spin diffusion at late times.

\begin{figure}
    \includegraphics{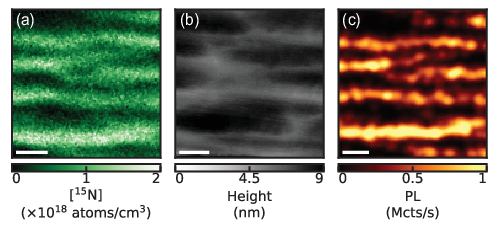}
    \caption{
    Co-localized materials characterization of sample L043 reveals the nitrogen incorporation during growth, together with common features shared by the surface topography and the NV distribution.
    (a) NanoSIMS image of the third doped layer (late in growth), showing inhomogeneous $^{15}$N incorporation (tracked via $^{12}$C$^{15}$N$^{-}$ ion counts) with striated features.
    (b) AFM image showing the post-growth surface topography, indicating that the surface features correlate with the local nitrogen density in (a).
    (c) PL image showing the presence of NVs along striations correlated to those in (a) and (b). All scale bars are \SI{1}{\micro\meter}.
    }
    \label{fig:Compare4Images}
\end{figure}

\textit{Engineering one-dimensional confinement.}---To engineer one-dimensional confinement of spin defects in diamond, we combine vertical confinement via nitrogen delta-doping~\cite{hughes_two-dimensional_2023,ohno_engineering_2012} with lateral confinement via preferential incorporation along atomic steps on a vicinal (100)-oriented diamond surface~\cite{meynell_engineering_2020}. 
Moreover, step bunches (aggregates of step edges) may form during growth~\cite{de_theije_effects_2000, frank_kinematic_1958, kandel_theory_1994, chayahara_effect_2004, achard_coupled_2007}, which both laterally confine and enhance nitrogen density [as depicted schematically in Fig.~\ref{fig:Introduction}(a)], enabling the realization of interacting 1D spin ensembles.
The quasi-1D incorporation of underlying nitrogen is also reflected in confocal photoluminescence images of NV centers [Fig.~\ref{fig:Introduction}(b)]

In this work, we characterize two grown samples (L043 and S011) consisting of isotopically purified diamond epilayers on miscut (100) substrates.
L043 is a three-layer $^{15}$N-doped sample, while S011 is a single-layer $^{14}$N-doped sample. 

We begin by characterizing the relationship between step bunching and defect incorporation.
Working with sample L043, we co-localize three distinct measurement modalities~\cite{sup_note}: (i) spatially resolved secondary ion mass spectrometry (nanoSIMS), which measures the local nitrogen concentration, (ii) atomic force microscopy (AFM), which measures the local post-growth surface topography, and (iii) confocal photoluminescence microscopy (PL), which measures the local NV density [Fig.~\ref{fig:Compare4Images}].
Notably, we observe common features across all three measurements.
The nanoSIMS data [Fig.~\ref{fig:Compare4Images}(a)] exhibit strong striations of increased nitrogen concentration, indicating that nitrogen preferentially incorporates into the observed topographic features that form during growth [Fig.~\ref{fig:Compare4Images}(b)]~\cite{sup_note}.
Moreover, these striations also closely mirror those seen in the confocal photoluminescence images [Fig.~\ref{fig:Compare4Images}(c)], which demonstrates that NV centers also preferentially incorporate in the same manner.

\begin{figure}
\includegraphics{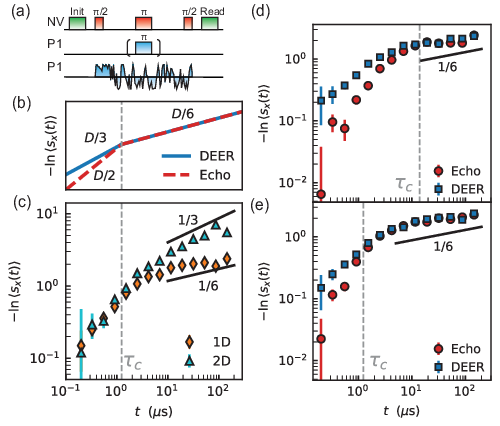}
\caption{(a) Pulse sequence for DEER-based extraction of P1 dimensionality. The parentheses around the P1 $\pi$-pulse indicate whether the measurement is a Hahn echo (no concurrent $\pi$-pulse) or DEER (concurrent $\pi$-pulse). The stochastic drive on the P1s (bottom line in sequence) is on for the entire duration of both measurements other than the $\pi$-pulse during DEER. The NV pulses correspond to laser (green) and microwave $\pi$- and $\pi/2$-pulses (red). (b) Idealized Hahn echo and DEER decay curves expected for disordered spin ensembles of dimensionality $D$ with a correlation time indicated by the dashed line. 
(c) Comparison of stochastically driven DEER decays on 1D and 2D regions of the sample, displaying a clear difference in stretch exponent in accordance with theory. Due to differences in density that change the absolute coherence level, the displayed 2D curve has been vertically shifted (multiplied by 2) for clarity. (d), (e) Full Hahn echo and DEER decoherence curves for $\tau_c = \SI{13.9}{\micro\second}$ (d) and $\SI{1.24}{\micro\second}$ (e), showing unambiguous one-dimensionality at late times regardless of the intermediate-time dynamics.}
\label{fig:dimensionality}
\end{figure}

\textit{Nanoscale probe of dimensionality.}---Taken together, the co-localized measurements in Fig.~\ref{fig:Compare4Images} establish the nanoscale mechanism that confines nitrogen laterally.
However, to establish the dimension of the spin ensemble, one requires a more local characterization method.
To this end, following the methodology developed in Ref.~\cite{davis_probing_2023}, we use the optically addressable NV centers to probe the magnetic noise generated by the dynamics of the optically dark, unpolarized P1 ensemble~\cite{milov_double_1997, kattnig_modeling_2013, salikhov_theory_1981, kutsovsky_electron_1990, lacelle_dipolar_1995, feldman_configurational_1996}.

Turning to single-layer sample S011, which also exhibits lateral confinement~\cite{sup_note}, we perform Hahn echo and double electron-electron resonance (DEER) measurements (which selectively probe the P1 dynamics) [Fig.~\ref{fig:dimensionality}(a)] and analyze the resulting NV decoherence curves. 
The decoherence curve $\langle s_x(t)\rangle$ encodes information about both the dimensionality and underlying fluctuations of the P1 spins.  
To constrain the spin dynamics for the purpose of extracting dimensionality, we impose a known noise profile on the P1 spins via a continuous phase-modulated drive~\cite{joos_protecting_2022} [Fig.~\ref{fig:dimensionality}(a)]. 
This ``stochastic'' drive forces each P1 spin to undergo Gauss-Markov dynamics with an autocorrelation function $\langle p^z_i(t)p^z_i(0)\rangle \sim \mathrm{e}^{-t/\tau_c}$; here, $p^z_i(t)$ is the $z$-spin operator for the $i$-th P1 spin in the ensemble, and $\tau_c$ is the correlation time, which can be experimentally tuned via the amplitude and bandwidth of the drive.
As depicted in Fig.~\ref{fig:dimensionality}(b), for a dilute probe spin (NV center) coupled to a positionally disordered spin bath (P1 centers) via long-range interactions (decaying with power-law $\alpha$), both Hahn echo and DEER decoherence profiles collapse at late times (i.e., $t >\tau_c$) to the same stretched exponential decay, $\langle s_x(t) \rangle \sim \mathrm{e}^{-(t/T_2)^{D/2\alpha}}$~\cite{davis_probing_2023}. 
For a 1D dipolar spin system, $D = 1$ and $\alpha = 3$, which produces a stretch exponent of $1/6$ [Fig.~\ref{fig:dimensionality}(b)]. 
We use the presence of a $1/6$ stretch exponent as evidence for one-dimensionality in the spin dynamics.

Figures~\ref{fig:dimensionality}(d) and~\ref{fig:dimensionality}(e) show Hahn echo and DEER measurements under stochastic drives that target P1 autocorrelation times of $\tau_c = \SI{13.9}{\micro\second}$ and $\tau_c = \SI{1.24}{\micro\second}$, respectively. 
We observe a slow decay characterized by a stretch exponent of $1/6$ at late times, confirming one-dimensionality.
Additionally, we find that Hahn echo and DEER profiles collapse at a time consistent with the expected $\tau_c$, confirming our control over P1 dynamics [Figs.~\ref{fig:dimensionality}(d) and~\ref{fig:dimensionality}(e)].

We note that the stretch exponent varies across the sample, consistent with the observation that S011 exhibits different degrees of striated photoluminescence in different regions of the sample. 
For comparison, we plot DEER measurements (again under a driven correlation time of $\tau_c = \SI{1.2}{\micro\second}$) from two exemplar regions in Fig.~\ref{fig:dimensionality}(c), one which exhibits a late-time stretch exponent of $1/6$ and one which exhibits a stretch exponent of $1/3$. 
We interpret the observed stretch exponents as revealing differences between the average nanoscale dipolar structure in these two regions, indicating locally 1D and 2D dimensions, respectively. 
Details about the difference between these two regions can be found in the Supplemental Material~\cite{sup_note}.
\begin{figure}
\includegraphics{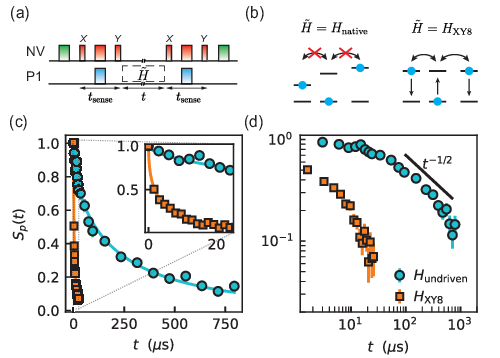}
\caption{(a) Pulse sequence for the correlation measurement. $t_{\text{sense}} \ll t$ is the DEER sensing time held constant for all measurements. 
Hamiltonian engineering is performed on the P1s to achieve an effective Hamiltonian $\tilde{H}$, and dynamics induced by this effective Hamiltonian are measured as a function of the evolution time $t$.
(b) Schematic of the P1 dynamics. 
On-site disorder in $H_\text{native}$ induces an energetic mismatch that suppresses P1-P1 flip-flops, while removal of that disorder in $H_\text{XY8}$ permits it. 
(c) Normalized correlation signal for $H_\text{native}$ (cyan circles) and $H_\text{XY8}$ (orange squares). 
Operationally, $H_\text{native}$ is measured without microwave pulses on the P1s. 
Inset: early-time decay of the autocorrelation to extract the decay timescale. 
Solid lines: stretched exponential fits of the decay for extracting decay timescales, where $H_\text{native}$ admits a decay timescale of $\SI{200}{\micro\second}$ while $H_\text{XY8}$ admits a decay timescale of $\SI{3}{\micro\second}$. 
(d) The same data as in (c), reproduced on a log-log plot.}
\label{fig:correlation}
\end{figure}

\textit{High-temperature, disordered dipolar dynamics in 1D.}---Having used stochastically driven P1 dynamics to reveal the 1D nature of the P1 spin system, we now consider the system's intrinsic dynamics driven by dipole-dipole XXZ interactions:
\begin{equation}
    H = \sum_i h_i p^z_i + \sum_{i < j} \frac{J_0}{r_{ij}^3} \left(\frac{1}{2}\left(p^x_i p^x_j + p^y_i p^y_j\right) - p^z_i p^z_j\right),
    \label{eq:native}
\end{equation}
and probe spin dynamics driven by these interactions. 
Here, $J_0 = 2\pi \times \SI{52}{\mega\hertz}$ and $r_{ij}$ is the distance between the $i$-th and $j$-th spins.
We note three features of this Hamiltonian: (i) there are ``random'' on-site fields, $h_i$, arising from the fact that our experiments only address a single spectral group (constituting 1/4 of the total P1 population), leaving off-resonant groups to generate disorder~\cite{sup_note}, (ii) the interactions obey a $1/r^3$ power-law dependence, and (iii) the total P1 $z$-magnetization, $\sum p_i^z$, is conserved.
These latter two facts, in combination with the completely unpolarized nature of the P1 bath, motivate the study of infinite-temperature spin transport in our dipolar system~\cite{wei_quantum_2022,jepsen_spin_2020}.

Experimentally, spin transport has been characterized in a variety of 1D platforms with long-range (typically power-law) interactions in the absence of positional disorder~\cite{joshi_observing_2022, peng_exploiting_2023}.
For sufficiently long-range couplings, transport becomes superdiffusive, and L\'evy flights have been predicted and observed~\cite{joshi_observing_2022, schuckert_nonlocal_2020}.
Dipolar interactions, by contrast, are sufficiently short-ranged to yield diffusion in 1D~\cite{kloss_spin_2019, schuckert_nonlocal_2020}. 
Yet, under strong positional disorder, the large distribution of interaction strengths could in principle modify this picture, and theoretical predictions range from diffusion to logarithmically slow relaxation and localization~\cite{yao2014many, zhao2025superspin, nandkishore_lifetimes_2021, stasiuk_disorder-induced_2025, selco_continuous-time_2026}.

Conventionally, in order to access a system's spin transport behavior, one measures the local spin survival probability, $S_p(t) \sim \langle p_i^z(t) p_i^z(0) \rangle$~\cite{zu_emergent_2021, zhao2025superspin}; however, with experimental access to only global observables, it is challenging to probe such local autocorrelation functions.
While prior experiments on solid-state spins have exploited the randomness of on-site fields to access local autocorrelations~\cite{peng_exploiting_2023, martin_controlling_2023}, here we demonstrate that correlation spectroscopy~\cite{rezai_probing_2025,dwyer_probing_2022,yu_engineering_2026} on optically dark, unpolarized spins exploits randomness in NV-dark spin dipolar couplings to achieve the same goal. 

Correlation spectroscopy consists of two temporally separated NV-P1 DEER-like pulse sequences [Fig.~\ref{fig:correlation}(a)], each of which imprints the instantaneous magnetic field $B_\text{P1}(t)$ produced by the P1 bath at the NV's location into the NV phase. 
This sequence enables a direct measurement of the autocorrelation function $\langle B_\text{P1}(t)B_\text{P1}(0) \rangle$~\cite{rezai_probing_2025,zhang_reporter-spin-assisted_2023,laraoui_high-resolution_2013,liu_surface_2022, yu_engineering_2026}.
After averaging across NV centers, the total correlation signal is composed of two terms: (i) the autocorrelation function $S_p(t)$, which is a probe of transport behavior, and (ii) cross-correlation functions between different P1s, which are not:
\begin{equation}
    \overline{\langle B_\text{P1}(t) B_\text{P1}(0) \rangle} = \overline{\left\langle \sum_{i} \frac{J_0^2}{r_i^6} p^z_i(t) p^z_i(0) \right\rangle} + \overline{\left\langle \sum_{i \neq j} \frac{J_0^2}{r_i^3 r_j^3} p^z_i(t) p^z_j(0) \right\rangle}.
\label{eq:autocorrelation}
\end{equation}
Here, $r_i$ is the distance between the probe NV and P1 spin $i$.

We emphasize that the expectation values in Eq.~(\ref{eq:autocorrelation}) contain two averages: an infinite-temperature average over P1 spin states $\langle \cdots \rangle$, and an average over NV-P1 positional configurations $\overline{\rule{0pt}{1ex}\cdots\rule{0pt}{1ex}}$ that arises from the random locations of the P1s relative to the NV probe spins. 
The infinite-temperature average ensures that the cross terms are zero at $t = 0$.
Moreover, even at non-zero times, the average over positional configurations ensures that the cross terms should be parametrically smaller than the local spin survival probability. 
In the End Matter, we analytically quantify these expectations, and provide numerical simulations that bound the norm of the second term to $\lesssim 2 \%$ of the first term for a 1D spin ensemble of our experimental density.
Thus, we take the correlation spectroscopy signal as a direct proxy for the local spin survival probability upon normalization, i.e.,
\begin{equation}
    \overline{\langle B_\text{P1}(t) B_\text{P1}(0) \rangle}\propto S_p(t).
\end{equation}

In Fig.~\ref{fig:correlation}, we compare two correlation spectroscopy measurements of P1 dynamics (on linear (c) and log (d) axes) under a native Hamiltonian and under an XY8 dynamical decoupling sequence. 
We make three observations.
First, the native spin transport dynamics occur on a significantly slower timescale than the typical dipolar interaction strength between P1 spins given our density estimates~\cite{sup_note}. 
The slow decay in native dynamics arises from the presence of strong disorder, which suppresses resonant flip-flop dynamics that are first-order in $J_0/r_{ij}^3$ and yields a spin-polarization transfer rate at second order, $\sim (J_0/r_{ij}^3)^2/|h_i - h_j|$.

Second, dynamical decoupling accelerates the decay of $S_p(t)$ by two orders of magnitude relative to native dynamics.
The accelerated decay arises because XY8 reduces the strength of on-site disorder~\cite{choi_robust_2020}, enabling the dipolar flip-flop dynamics to become more resonant, as schematically shown in Fig.~\ref{fig:correlation}(b). 
We emphasize that this acceleration establishes the dominance of dipolar interactions in our system and justifies the Hamiltonian description of Eq.~(\ref{eq:native}).

Third, the survival probability under the native Hamiltonian exhibits a window consistent with diffusive behavior between $\SI{50}{\micro\second} < t < \SI{400}{\micro\second}$ [Fig.~\ref{fig:correlation}(d)], where $S_p(t) \sim t^{-\nu}$ and $\nu=1/2$ for 1D diffusion.
At late times, $S_p(t)$ decays more quickly than $t^{-1/2}$. 
We attribute this regime to phonon-induced depolarization, which causes an exponential decay with a timescale $\sim 1$~ms. 
Therefore, further measurements at low temperature to access more decades of time are required to more robustly test theoretical models of transport.  

We note that under the XY8-decoupled Hamiltonian, evaluating late-time transport dynamics is challenging since the correlation spectroscopy reaches its noise floor at $t \sim \SI{10}{\micro\second}$. 
To this end, it would be intriguing to undertake further measurements in order to distinguish the role of positional disorder relative to on-site disorder.

\textit{Outlook.}---Looking forward, our work opens the door to a number of intriguing future directions. 
First, leveraging surface topography in a more controlled manner---for instance, via patterned etching prior to growth---opens new avenues for creating custom spin geometries in diamond~\cite{postelnicu_harnessingdiamond_2026, gotze_preferential_2022, lang_controlled_2024, achard_chemical_2020, hayashi_selective_2026}.
Second, by using optically polarizable NV centers, state preparation with spatial control becomes possible. 
P1 centers may be converted to NV centers via electron irradiation, or the incorporation of grown-in NV centers may be boosted, either of which may enable reaching the 1D dipolar-interacting limit with NV centers themselves.
Third, spin transport plays an important role in facilitating polarization transfer between optically initialized spins (such as NV centers) and dark spins in the environment (such as P1 centers or surface spins).
Understanding the interplay between disorder and interaction~\cite{javed_magic-novel_2025, jain_off-resonance_2017} is essential to realizing proposed schemes that deploy dark spins as resources for quantum sensing\cite{cappellaro_environment-assisted_2012, goldstein_environment-assisted_2011, cooper_environment-assisted_2019, zhang_reporter-spin-assisted_2023, sushkov_magnetic_2014}. 
Finally, it may be possible to explore the transition between one- and two-dimensional, disordered dipolar spin transport via the use of magnetic field gradients~\cite{put2025collectivemanybodydynamicssolidstate}.

\emph{Acknowledgments.}---We thank Yu-Xin Wang, Bingtian Ye, Haoyang Gao, Yi Zhao, Joel Moore, Emily Davis, and Xiaofei Yu for helpful discussions.
We thank Christine Jilly (ORCID: 0000-0003-4879-8739) for instrumental help with conducting nanoSIMS measurements.
This work was supported by the U.S. Department of Energy via BES grant DE-SC0019241 (materials synthesis and characterization), the Army Research Office through the MURI program grant no.~W911NF-20-1-0136 (theoretical studies), and the U.S. DOE BES grant DE-SC0026470 (numerical simulations).
We acknowledge the use of shared facilities of the UCSB Quantum Foundry through the Q-AMASE-i program (NSF DMR-1906325), the UCSB MRSEC (NSF DMR-2308708), and the Quantum Structures Facility within the UCSB California NanoSystems Institute. 
A portion of this work was performed in the UCSB Nanofabrication Facility, an open access laboratory.
Part of this work was performed at nano@stanford RRID:SCR\_026695. 
The computations in this paper were run on the FASRC FASSE cluster supported by the FAS Division of Science Research Computing Group at Harvard University.
S. P. acknowledges support from the Department of Defense (DoD) through the National Defense Science and Engineering Graduate (NDSEG) Fellowship Program.
S. A. M. acknowledges support from Canada NSERC (AID 516704-2018) and the UCSB Quantum Foundry. 
L. B. H. W. acknowledges support from the NSF Graduate Research Fellowship Program (DGE 2139319) and the UCSB Quantum Foundry.
A. C. B. J.  and N. Y. Y. acknowledge support from the NSF QLCI program through grant number OMA-2016245.

\bibliography{1D_NanoSIMS_GSD.bib}

@misc{sup_note,
	note = {See Supplemental Material at [URL will be inserted by publisher] for details on sample characterization, measurement, and theoretical derivation, which includes Ref.~[73].}
}

@article{eichhorn_optimizing_2019,
	title = {Optimizing the formation of depth-confined nitrogen vacancy center spin ensembles in diamond for quantum sensing},
	volume = {3},
	issn = {2475-9953},
	url = {https://link.aps.org/doi/10.1103/PhysRevMaterials.3.113802},
	doi = {10.1103/PhysRevMaterials.3.113802},
	language = {},
	number = {11},
	urldate = {2026-09-01},
	journal = {Physical Review Materials},
	author = {Eichhorn, Tim R. and McLellan, Claire A. and Bleszynski Jayich, Ania C.},
	month = nov,
	year = {2019},
	pages = {113802},
}

@article{sushkov_magnetic_2014,
	title = {Magnetic {Resonance} {Detection} of {Individual} {Proton} {Spins} {Using} {Quantum} {Reporters}},
	volume = {113},
	copyright = {http://link.aps.org/licenses/aps-default-license},
	issn = {0031-9007, 1079-7114},
	url = {https://link.aps.org/doi/10.1103/PhysRevLett.113.197601},
	doi = {10.1103/PhysRevLett.113.197601},
	language = {},
	number = {19},
	urldate = {2026-09-01},
	journal = {Physical Review Letters},
	author = {Sushkov, A.O. and Lovchinsky, I. and Chisholm, N. and Walsworth, R.L. and Park, H. and Lukin, M.D.},
	month = nov,
	year = {2014},
	pages = {197601},
}

@article{achard_coupled_2007,
	title = {Coupled effect of nitrogen addition and surface temperature on the morphology and the kinetics of thick {CVD} diamond single crystals},
	volume = {16},
	copyright = {https://www.elsevier.com/tdm/userlicense/1.0/},
	issn = {09259635},
	url = {https://linkinghub.elsevier.com/retrieve/pii/S0925963506003220},
	doi = {10.1016/j.diamond.2006.09.012},
	language = {},
	number = {4-7},
	urldate = {2026-08-25},
	journal = {Diamond and Related Materials},
	author = {Achard, J. and Silva, F. and Brinza, O. and Tallaire, A. and Gicquel, A.},
	month = apr,
	year = {2007},
	pages = {685--689},
}

@article{chayahara_effect_2004,
	title = {The effect of nitrogen addition during high-rate homoepitaxial growth of diamond by microwave plasma {CVD}},
	volume = {13},
	copyright = {https://www.elsevier.com/tdm/userlicense/1.0/},
	issn = {09259635},
	url = {https://linkinghub.elsevier.com/retrieve/pii/S0925963504002511},
	doi = {10.1016/j.diamond.2004.07.007},
	language = {},
	number = {11-12},
	urldate = {2026-08-25},
	journal = {Diamond and Related Materials},
	author = {Chayahara, A. and Mokuno, Y. and Horino, Y. and Takasu, Y. and Kato, H. and Yoshikawa, H. and Fujimori, N.},
	month = nov,
	year = {2004},
	pages = {1954--1958},
}

@article{wang_manipulating_2023,
	title = {Manipulating solid-state spin concentration through charge transport},
	volume = {120},
	issn = {0027-8424, 1091-6490},
	url = {https://pnas.org/doi/10.1073/pnas.2305621120},
	doi = {10.1073/pnas.2305621120},
	language = {},
	number = {32},
	urldate = {2026-07-31},
	journal = {Proceedings of the National Academy of Sciences},
	author = {Wang, Guoqing and Li, Changhao and Tang, Hao and Li, Boning and Madonini, Francesca and Alsallom, Faisal F. and Calvin Sun, Won Kyu and Peng, Pai and Villa, Federica and Li, Ju and Cappellaro, Paola},
	month = aug,
	year = {2023},
	pages = {e2305621120},
}

@article{monge_beyond_2025,
	title = {Beyond ensemble averaging: {Parallelized} single-shot readout of hole capture in diamond},
	copyright = {Creative Commons Attribution 4.0 International},
	shorttitle = {Beyond ensemble averaging},
	url = {https://arxiv.org/abs/2507.11722},
	journal = {arXiv preprint},
	author = {Monge, Richard and Nakamura, Yuki and Bach, Olaf and Shao, Jason and Lozovoi, Artur and Wood, Alexander A. and Sasaki, Kento and Kobayashi, Kensuke and Delord, Tom and Meriles, Carlos A.},
	year = {2025},
}

@article{mukherjee_influence_2024,
	title = {Influence of disordered and anisotropic interactions on relaxation dynamics and propagation of correlations in tweezer arrays of {Rydberg} dipoles},
	volume = {110},
	issn = {2469-9926, 2469-9934},
	url = {https://link.aps.org/doi/10.1103/PhysRevA.110.053320},
	doi = {10.1103/PhysRevA.110.053320},
	language = {},
	number = {5},
	urldate = {2026-07-27},
	journal = {Physical Review A},
	author = {Mukherjee, K. and Biedermann, G. W. and Lewis-Swan, R. J.},
	month = nov,
	year = {2024},
	pages = {053320}
}

@article{hazzard_many-body_2014,
	title = {Many-{Body} {Dynamics} of {Dipolar} {Molecules} in an {Optical} {Lattice}},
	volume = {113},
	copyright = {http://link.aps.org/licenses/aps-default-license},
	issn = {0031-9007, 1079-7114},
	url = {https://link.aps.org/doi/10.1103/PhysRevLett.113.195302},
	doi = {10.1103/PhysRevLett.113.195302},
	language = {},
	number = {19},
	urldate = {2026-07-27},
	journal = {Physical Review Letters},
	author = {Hazzard, Kaden R. A. and Gadway, Bryce and Foss-Feig, Michael and Yan, Bo and Moses, Steven A. and Covey, Jacob P. and Yao, Norman Y. and Lukin, Mikhail D. and Ye, Jun and Jin, Deborah S. and Rey, Ana Maria},
	month = nov,
	year = {2014},
	pages = {195302}
}

@article{put2025collectivemanybodydynamicssolidstate,
      title={Collective many-body dynamics in a solid-state quantum sensor controlled through nanoscale magnetic gradients}, 
      author={Piotr Put and Nathaniel T. Leitao and Haoyang Gao and Christina Spaegele and Oksana Makarova and Lillian B. Hughes Wyatt and Andrew C. Maccabe and Matthew Mammen and Bartholomeus Machielse and Hengyun Zhou and Szymon Pustelny and Ania C. Bleszynski Jayich and Federico Capasso and Leigh S. Martin and Hongkun Park and Mikhail D. Lukin},
      year={2025},
      journal={arXiv preprint},
      url={https://arxiv.org/abs/2506.11920}, 
}

@article{frank_kinematic_1958,
journal = {Growth and Perfection of Crystals},
title = {On the kinematic theory of crystal growth and dissolution processes},
volume = {411},
year = {1958},
author = {FRANK F. C.},
}

@article{ammerlaan_reorientation_1981,
	title = {Reorientation of {Nitrogen} in {Type}-{I}b {Diamond} by {Thermal} {Excitation} and {Tunneling}},
	volume = {47},
	copyright = {http://link.aps.org/licenses/aps-default-license},
	issn = {0031-9007},
	url = {https://link.aps.org/doi/10.1103/PhysRevLett.47.954},
	doi = {10.1103/PhysRevLett.47.954},
	language = {},
	number = {13},
	urldate = {2026-07-30},
	journal = {Physical Review Letters},
	author = {Ammerlaan, C. A. J. and Burgemeister, E. A.},
	month = sep,
	year = {1981},
	pages = {954--957}
}

@article{xiao_proposal_2016,
	title = {Proposal for observing dynamic {Jahn}-{Teller} effect by single solid-state defects},
	volume = {18},
	issn = {1367-2630},
	url = {https://iopscience.iop.org/article/10.1088/1367-2630/18/10/103022},
	doi = {10.1088/1367-2630/18/10/103022},
	number = {10},
	urldate = {2026-07-30},
	journal = {New Journal of Physics},
	author = {Xiao, Xing and Zhao, Nan},
	month = oct,
	year = {2016},
	pages = {103022}
}

@article{grinolds_subnanometre_2014,
	title = {Subnanometre resolution in three-dimensional magnetic resonance imaging of individual dark spins},
	volume = {9},
	copyright = {http://www.springer.com/tdm},
	issn = {1748-3387, 1748-3395},
	url = {https://www.nature.com/articles/nnano.2014.30},
	doi = {10.1038/nnano.2014.30},
	language = {},
	number = {4},
	urldate = {2026-07-30},
	journal = {Nature Nanotechnology},
	author = {Grinolds, M. S. and Warner, M. and De Greve, K. and Dovzhenko, Y. and Thiel, L. and Walsworth, R. L. and Hong, S. and Maletinsky, P. and Yacoby, A.},
	month = apr,
	year = {2014},
	pages = {279--284}
}

@article{feldman_configurational_1996,
	title = {Configurational averaging of dipolar interactions in magnetically diluted spin networks},
	volume = {104},
	issn = {0021-9606, 1089-7690},
	url = {https://pubs.aip.org/jcp/article/104/5/2000/478641/Configurational-averaging-of-dipolar-interactions},
	doi = {10.1063/1.470956},
	language = {},
	number = {5},
	urldate = {2024-04-26},
	journal = {The Journal of Chemical Physics},
	author = {Fel'dman, Edward B. and Lacelle, Serge},
	month = feb,
	year = {1996},
	pages = {2000--2009}
}

@article{gao_signal_2025,
	title = {Signal amplification in a solid-state sensor through asymmetric many-body echo},
	volume = {646},
	issn = {0028-0836, 1476-4687},
	url = {https://www.nature.com/articles/s41586-025-09452-7},
	doi = {10.1038/s41586-025-09452-7},
	language = {},
	number = {8083},
	urldate = {2025-11-05},
	journal = {Nature},
	author = {Gao, Haoyang and Martin, Leigh S. and Hughes, Lillian B. and Leitao, Nathaniel T. and Put, Piotr and Zhou, Hengyun and Koyluoglu, Nazli U. and Meynell, Simon A. and Jayich, Ania C. Bleszynski and Park, Hongkun and Lukin, Mikhail D.},
	month = oct,
	year = {2025},
	pages = {68--73}
}

@article{lacelle_dipolar_1995,
	title = {Dipolar interactions in magnetically very diluted spin networks},
	volume = {102},
	issn = {0021-9606, 1089-7690},
	url = {https://pubs.aip.org/jcp/article/102/2/947/479877/Dipolar-interactions-in-magnetically-very-diluted},
	doi = {10.1063/1.469162},
	language = {},
	number = {2},
	urldate = {2024-04-26},
	journal = {The Journal of Chemical Physics},
	author = {Lacelle, Serge and Tremblay, Luc},
	month = jan,
	year = {1995},
	pages = {947--955}
}

@article{kutsovsky_electron_1990,
	title = {Electron spin echo as a tool for investigation of surface structure of finely dispersed fractal solids},
	volume = {42},
	copyright = {http://www.springer.com/tdm},
	issn = {0133-1736, 1588-2837},
	url = {http://link.springer.com/10.1007/BF02137612},
	doi = {10.1007/BF02137612},
	language = {},
	number = {1},
	urldate = {2026-07-20},
	journal = {Reaction Kinetics \& Catalysis Letters},
	author = {Kutsovsky, Y. E. and Mariasov, A. G. and Aristov, Y. I. and Parmon, V. N.},
	month = jul,
	year = {1990},
	pages = {19--24}
}

@article{salikhov_theory_1981,
	title = {The theory of electron spin-echo signal decay resulting from dipole-dipole interactions between paramagnetic centers in solids},
	volume = {42},
	issn = {00222364},
	url = {https://linkinghub.elsevier.com/retrieve/pii/002223648190216X},
	doi = {10.1016/0022-2364(81)90216-X},
	language = {},
	number = {2},
	urldate = {2026-07-26},
	journal = {Journal of Magnetic Resonance (1969)},
	author = {Salikhov, K.M and Dzuba, S.A and Raitsimring, A.M},
	month = feb,
	year = {1981},
	pages = {255--276}
}

@article{kattnig_modeling_2013,
	title = {Modeling {Excluded} {Volume} {Effects} for the {Faithful} {Description} of the {Background} {Signal} in {Double} {Electron}-{Electron} {Resonance}},
	volume = {117},
	issn = {1520-6106, 1520-5207},
	url = {https://pubs.acs.org/doi/10.1021/jp408338q},
	doi = {10.1021/jp408338q},
	language = {},
	number = {51},
	urldate = {2026-07-26},
	journal = {The Journal of Physical Chemistry B},
	author = {Kattnig, Daniel R. and Reichenwallner, Jörg and Hinderberger, Dariush},
	month = dec,
	year = {2013},
	pages = {16542--16557}
}

@article{milov_double_1997,
	title = {Double electron-electron resonance in electron spin echo: {Conformations} of spin-labeled poly-4-vinilpyridine in glassy solutions},
	volume = {12},
	copyright = {http://www.springer.com/tdm},
	issn = {0937-9347, 1613-7507},
	shorttitle = {Double electron-electron resonance in electron spin echo},
	url = {http://link.springer.com/10.1007/BF03164129},
	doi = {10.1007/BF03164129},
	language = {},
	number = {4},
	urldate = {2026-07-26},
	journal = {Applied Magnetic Resonance},
	author = {Milov, A. D. and Tsvetkov, Yu. D.},
	month = may,
	year = {1997},
	pages = {495--504}
}

@article{yu_engineering_2026,
	title = {Engineering diamond interfaces free of dark spins},
	volume = {25},
	issn = {2331-7019},
	url = {https://link.aps.org/doi/10.1103/3j8r-2kt6},
	doi = {10.1103/3j8r-2kt6},
	language = {},
	number = {3},
	urldate = {2026-07-16},
	journal = {Physical Review Applied},
	author = {Yu, Xiaofei and Villafranca, Evan J. and Wang, Stella and Jones, Jessica C. and Xie, Mouzhe and Nagura, Jonah and Chi-Dur{\'a}n, Ignacio and Delegan, Nazar and Martinson, Alex B.F. and Flatt{\'e}, Michael E. and Candido, Denis R. and Galli, Giulia and Maurer, Peter C.},
	month = mar,
	year = {2026},
	pages = {034006}
}

@article{javed_magic-novel_2025,
	title = {Magic-{NOVEL}: {Suppressing} electron-electron coupling effects in pulsed {DNP}},
	volume = {162},
	issn = {0021-9606, 1089-7690},
	shorttitle = {Magic-{NOVEL}},
	url = {https://pubs.aip.org/jcp/article/162/1/014202/3329051/Magic-NOVEL-Suppressing-electron-electron-coupling},
	doi = {10.1063/5.0241288},
	language = {},
	number = {1},
	urldate = {2026-07-04},
	journal = {The Journal of Chemical Physics},
	author = {Javed, Amaria and Ghazi, Marwa Yaser and SubbaRao Redrouthu, Venkata and Equbal, Asif},
	month = jan,
	year = {2025},
	pages = {014202},
}

@article{jain_off-resonance_2017,
	title = {Off-resonance {NOVEL}},
	volume = {147},
	issn = {0021-9606, 1089-7690},
	url = {https://pubs.aip.org/jcp/article/147/16/164201/76946/Off-resonance-NOVEL},
	doi = {10.1063/1.5000528},
	language = {},
	number = {16},
	urldate = {2026-07-04},
	journal = {The Journal of Chemical Physics},
	author = {Jain, Sheetal K. and Mathies, Guinevere and Griffin, Robert G.},
	month = oct,
	year = {2017},
	pages = {164201},
}

@article{dwyer_probing_2022,
	title = {Probing {Spin} {Dynamics} on {Diamond} {Surfaces} {Using} a {Single} {Quantum} {Sensor}},
	volume = {3},
	issn = {2691-3399},
	url = {https://link.aps.org/doi/10.1103/PRXQuantum.3.040328},
	doi = {10.1103/PRXQuantum.3.040328},
	language = {},
	number = {4},
	urldate = {2026-07-04},
	journal = {PRX Quantum},
	author = {Dwyer, Bo L. and Rodgers, Lila V.H. and Urbach, Elana K. and Bluvstein, Dolev and Sangtawesin, Sorawis and Zhou, Hengyun and Nassab, Yahia and Fitzpatrick, Mattias and Yuan, Zhiyang and De Greve, Kristiaan and Peterson, Eric L. and Knowles, Helena and Sumarac, Tamara and Chou, Jyh-Pin and Gali, Adam and Dobrovitski, V.V. and Lukin, Mikhail D. and De Leon, Nathalie P.},
	month = dec,
	year = {2022},
	pages = {040328},
}

@article{cooper_environment-assisted_2019,
	title = {Environment-assisted {Quantum}-enhanced {Sensing} with {Electronic} {Spins} in {Diamond}},
	volume = {12},
	issn = {2331-7019},
	url = {https://link.aps.org/doi/10.1103/PhysRevApplied.12.044047},
	doi = {10.1103/PhysRevApplied.12.044047},
	language = {},
	number = {4},
	urldate = {2026-07-04},
	journal = {Physical Review Applied},
	author = {Cooper, Alexandre and Sun, Won Kyu Calvin and Jaskula, Jean-Christophe and Cappellaro, Paola},
	month = oct,
	year = {2019},
	pages = {044047},
}

@article{goldstein_environment-assisted_2011,
	title = {Environment-{Assisted} {Precision} {Measurement}},
	volume = {106},
	copyright = {http://link.aps.org/licenses/aps-default-license},
	issn = {0031-9007, 1079-7114},
	url = {https://link.aps.org/doi/10.1103/PhysRevLett.106.140502},
	doi = {10.1103/PhysRevLett.106.140502},
	language = {},
	number = {14},
	urldate = {2026-07-04},
	journal = {Physical Review Letters},
	author = {Goldstein, G. and Cappellaro, P. and Maze, J. R. and Hodges, J. S. and Jiang, L. and S{\o}rensen, A. S. and Lukin, M. D.},
	month = apr,
	year = {2011},
	pages = {140502},
}

@article{cappellaro_environment-assisted_2012,
	title = {Environment-assisted metrology with spin qubits},
	volume = {85},
	copyright = {http://link.aps.org/licenses/aps-default-license},
	issn = {1050-2947, 1094-1622},
	url = {https://link.aps.org/doi/10.1103/PhysRevA.85.032336},
	doi = {10.1103/PhysRevA.85.032336},
	language = {},
	number = {3},
	urldate = {2026-07-04},
	journal = {Physical Review A},
	author = {Cappellaro, P. and Goldstein, G. and Hodges, J. S. and Jiang, L. and Maze, J. R. and S{\o}rensen, A. S. and Lukin, M. D.},
	month = mar,
	year = {2012},
	pages = {032336},
}

@article{rezai_probing_2025,
	title = {Probing {Dynamics} of a {Two}-{Dimensional} {Dipolar} {Spin} {Ensemble} {Using} {Single} {Qubit} {Sensor}},
	volume = {134},
	issn = {0031-9007, 1079-7114},
	url = {https://link.aps.org/doi/10.1103/PhysRevLett.134.050801},
	doi = {10.1103/PhysRevLett.134.050801},
	language = {},
	number = {5},
	urldate = {2026-07-04},
	journal = {Physical Review Letters},
	author = {Rezai, Kristine and Choi, Soonwon and Lukin, Mikhail D. and Sushkov, Alexander O.},
	month = feb,
	year = {2025},
	pages = {050801},
}

@article{peng_exploiting_2023,
	title = {Exploiting disorder to probe spin and energy hydrodynamics},
	volume = {19},
	issn = {1745-2473, 1745-2481},
	url = {https://www.nature.com/articles/s41567-023-02024-4},
	doi = {10.1038/s41567-023-02024-4},
	language = {},
	number = {7},
	urldate = {2026-07-04},
	journal = {Nature Physics},
	author = {Peng, Pai and Ye, Bingtian and Yao, Norman Y. and Cappellaro, Paola},
	month = jul,
	year = {2023},
	pages = {1027--1032},
}

@article{martin_controlling_2023,
	title = {Controlling {Local} {Thermalization} {Dynamics} in a {Floquet}-{Engineered} {Dipolar} {Ensemble}},
	volume = {130},
	issn = {0031-9007, 1079-7114},
	url = {https://link.aps.org/doi/10.1103/PhysRevLett.130.210403},
	doi = {10.1103/PhysRevLett.130.210403},
	language = {},
	number = {21},
	urldate = {2026-07-04},
	journal = {Physical Review Letters},
	author = {Martin, Leigh S. and Zhou, Hengyun and Leitao, Nathaniel T. and Maskara, Nishad and Makarova, Oksana and Gao, Haoyang and Zhu, Qian-Ze and Park, Mincheol and Tyler, Matthew and Park, Hongkun and Choi, Soonwon and Lukin, Mikhail D.},
	month = may,
	year = {2023},
	pages = {210403},
}

@article{zu_emergent_2021,
	title = {Emergent hydrodynamics in a strongly interacting dipolar spin ensemble},
	volume = {597},
	issn = {0028-0836, 1476-4687},
	url = {https://www.nature.com/articles/s41586-021-03763-1},
	doi = {10.1038/s41586-021-03763-1},
	language = {},
	number = {7874},
	urldate = {2026-07-04},
	journal = {Nature},
	author = {Zu, C. and Machado, F. and Ye, B. and Choi, S. and Kobrin, B. and Mittiga, T. and Hsieh, S. and Bhattacharyya, P. and Markham, M. and Twitchen, D. and Jarmola, A. and Budker, D. and Laumann, C. R. and Moore, J. E. and Yao, N. Y.},
	month = sep,
	year = {2021},
	pages = {45--50},
}

@article{joos_protecting_2022,
	title = {Protecting qubit coherence by spectrally engineered driving of the spin environment},
	volume = {8},
	issn = {2056-6387},
	url = {https://www.nature.com/articles/s41534-022-00560-0},
	doi = {10.1038/s41534-022-00560-0},
	language = {},
	number = {1},
	urldate = {2026-07-03},
	journal = {npj Quantum Information},
	author = {Joos, Maxime and Bluvstein, Dolev and Lyu, Yuanqi and Weld, David and Bleszynski Jayich, Ania},
	month = apr,
	year = {2022},
	pages = {47},
}

@article{affleck_quantum_1989,
	title = {Quantum spin chains and the {Haldane} gap},
	volume = {1},
	issn = {0953-8984, 1361-648X},
	url = {https://iopscience.iop.org/article/10.1088/0953-8984/1/19/001},
	doi = {10.1088/0953-8984/1/19/001},
	number = {19},
	urldate = {2026-07-03},
	journal = {Journal of Physics: Condensed Matter},
	author = {Affleck, I},
	month = may,
	year = {1989},
	pages = {3047--3072},
}

@article{de_theije_effects_2000,
	title = {Effects of nitrogen impurities on the {CVD} growth of diamond: step bunching in theory and experiment},
	volume = {9},
	copyright = {https://www.elsevier.com/tdm/userlicense/1.0/},
	issn = {09259635},
	shorttitle = {Effects of nitrogen impurities on the {CVD} growth of diamond},
	url = {https://linkinghub.elsevier.com/retrieve/pii/S0925963500002612},
	doi = {10.1016/S0925-9635(00)00261-2},
	language = {},
	number = {8},
	urldate = {2026-07-03},
	journal = {Diamond and Related Materials},
	author = {De Theije, F.K. and Schermer, J.J. and Van Enckevort, W.J.P.},
	month = aug,
	year = {2000},
	pages = {1439--1449},
}

@article{kandel_theory_1994,
	title = {Theory of impurity-induced step bunching},
	volume = {49},
	copyright = {http://link.aps.org/licenses/aps-default-license},
	issn = {0163-1829, 1095-3795},
	url = {https://link.aps.org/doi/10.1103/PhysRevB.49.5554},
	doi = {10.1103/PhysRevB.49.5554},
	language = {},
	number = {8},
	urldate = {2026-07-03},
	journal = {Physical Review B},
	author = {Kandel, Daniel and Weeks, John D.},
	month = feb,
	year = {1994},
	pages = {5554--5564},
}

@article{ohno_engineering_2012,
	title = {Engineering shallow spins in diamond with nitrogen delta-doping},
	volume = {101},
	issn = {0003-6951, 1077-3118},
	url = {https://pubs.aip.org/aip/apl/article/112025},
	doi = {10.1063/1.4748280},
	language = {},
	number = {8},
	urldate = {2026-07-03},
	journal = {Applied Physics Letters},
	author = {Ohno, Kenichi and Joseph Heremans, F. and Bassett, Lee C. and Myers, Bryan A. and Toyli, David M. and Bleszynski Jayich, Ania C. and Palmstr{\o}m, Christopher J. and Awschalom, David D.},
	month = aug,
	year = {2012},
	pages = {082413},
}

@article{meynell_engineering_2020,
	title = {Engineering quantum-coherent defects: {The} role of substrate miscut in chemical vapor deposition diamond growth},
	volume = {117},
	issn = {0003-6951, 1077-3118},
	shorttitle = {Engineering quantum-coherent defects},
	url = {https://pubs.aip.org/aip/apl/article/38680},
	doi = {10.1063/5.0029715},
	language = {},
	number = {19},
	urldate = {2026-07-03},
	journal = {Applied Physics Letters},
	author = {Meynell, Simon A. and McLellan, Claire A. and Hughes, Lillian B. and Wang, Wenbo and Mates, Tom E. and Mukherjee, Kunal and Bleszynski Jayich, Ania C.},
	month = nov,
	year = {2020},
	pages = {194001},
}

@article{hughes_two-dimensional_2023,
	title = {Two-dimensional spin systems in {PECVD}-grown diamond with tunable density and long coherence for enhanced quantum sensing and simulation},
	volume = {11},
	issn = {2166-532X},
	url = {https://pubs.aip.org/apm/article/11/2/021101/2870857/Two-dimensional-spin-systems-in-PECVD-grown},
	doi = {10.1063/5.0133501},
	language = {},
	number = {2},
	urldate = {2026-07-03},
	journal = {APL Materials},
	author = {Hughes, Lillian B. and Zhang, Zhiran and Jin, Chang and Meynell, Simon A. and Ye, Bingtian and Wu, Weijie and Wang, Zilin and Davis, Emily J. and Mates, Thomas E. and Yao, Norman Y. and Mukherjee, Kunal and Bleszynski Jayich, Ania C.},
	month = feb,
	year = {2023},
	pages = {021101},
}

@article{gong_coherent_2023,
	title = {Coherent dynamics of strongly interacting electronic spin defects in hexagonal boron nitride},
	volume = {14},
	issn = {2041-1723},
	url = {https://www.nature.com/articles/s41467-023-39115-y},
	doi = {10.1038/s41467-023-39115-y},
	language = {},
	number = {1},
	urldate = {2026-07-03},
	journal = {Nature Communications},
	author = {Gong, Ruotian and He, Guanghui and Gao, Xingyu and Ju, Peng and Liu, Zhongyuan and Ye, Bingtian and Henriksen, Erik A. and Li, Tongcang and Zu, Chong},
	month = jun,
	year = {2023},
	pages = {3299},
}

@article{davis_probing_2023,
	title = {Probing many-body dynamics in a two-dimensional dipolar spin ensemble},
	volume = {19},
	issn = {1745-2473, 1745-2481},
	url = {https://www.nature.com/articles/s41567-023-01944-5},
	doi = {10.1038/s41567-023-01944-5},
	language = {},
	number = {6},
	urldate = {2026-07-03},
	journal = {Nature Physics},
	author = {Davis, E. J. and Ye, B. and Machado, F. and Meynell, S. A. and Wu, W. and Mittiga, T. and Schenken, W. and Joos, M. and Kobrin, B. and Lyu, Y. and Wang, Z. and Bluvstein, D. and Choi, S. and Zu, C. and Jayich, A. C. Bleszynski and Yao, N. Y.},
	month = jun,
	year = {2023},
	pages = {836--844},
}

@article{choi_colloquium_2019,
	title = {\textit{{Colloquium}}: {Atomic} spin chains on surfaces},
	volume = {91},
	issn = {0034-6861, 1539-0756},
	shorttitle = {\textit{{Colloquium}}},
	url = {https://link.aps.org/doi/10.1103/RevModPhys.91.041001},
	doi = {10.1103/RevModPhys.91.041001},
	language = {},
	number = {4},
	urldate = {2026-07-03},
	journal = {Reviews of Modern Physics},
	author = {Choi, Deung-Jang and Lorente, Nicolas and Wiebe, Jens and Von Bergmann, Kirsten and Otte, Alexander F. and Heinrich, Andreas J.},
	month = oct,
	year = {2019},
	pages = {041001},
}

@article{li_imaging_2024,
	title = {Imaging tunable {Luttinger} liquid systems in van der {Waals} heterostructures},
	volume = {631},
	issn = {0028-0836, 1476-4687},
	url = {https://www.nature.com/articles/s41586-024-07596-6},
	doi = {10.1038/s41586-024-07596-6},
	language = {},
	number = {8022},
	urldate = {2026-07-03},
	journal = {Nature},
	author = {Li, Hongyuan and Xiang, Ziyu and Wang, Tianle and Naik, Mit H. and Kim, Woochang and Nie, Jiahui and Li, Shiyu and Ge, Zhehao and He, Zehao and Ou, Yunbo and Banerjee, Rounak and Taniguchi, Takashi and Watanabe, Kenji and Tongay, Sefaattin and Zettl, Alex and Louie, Steven G. and Zaletel, Michael P. and Crommie, Michael F. and Wang, Feng},
	month = jul,
	year = {2024},
	pages = {765--770},
}

@article{scheie_detection_2021,
	title = {Detection of {Kardar}-{Parisi}-{Zhang} hydrodynamics in a quantum {Heisenberg} spin-1/2 chain},
	volume = {17},
	issn = {1745-2473, 1745-2481},
	url = {https://www.nature.com/articles/s41567-021-01191-6},
	doi = {10.1038/s41567-021-01191-6},
	language = {},
	number = {6},
	urldate = {2026-07-03},
	journal = {Nature Physics},
	author = {Scheie, A. and Sherman, N. E. and Dupont, M. and Nagler, S. E. and Stone, M. B. and Granroth, G. E. and Moore, J. E. and Tennant, D. A.},
	month = jun,
	year = {2021},
	pages = {726--730},
}

@article{hayashi_nmr_1975,
	title = {{NMR} {Line} {Width} in {One}-{Dimensional} {Antiferromagnet} {KCuF}$_{\textrm{3}}$},
	volume = {38},
	issn = {0031-9015, 1347-4073},
	url = {https://journals.jps.jp/doi/10.1143/JPSJ.38.695},
	doi = {10.1143/JPSJ.38.695},
	language = {},
	number = {3},
	urldate = {2026-07-03},
	journal = {Journal of the Physical Society of Japan},
	author = {Hayashi, Hatsuo and Hirakawa, Kazuyoshi},
	month = mar,
	year = {1975},
	pages = {695--700},
}

@article{wei_quantum_2022,
	title = {Quantum gas microscopy of {Kardar}-{Parisi}-{Zhang} superdiffusion},
	volume = {376},
	issn = {0036-8075, 1095-9203},
	url = {https://www.science.org/doi/10.1126/science.abk2397},
	doi = {10.1126/science.abk2397},
	language = {},
	number = {6594},
	urldate = {2026-07-03},
	journal = {Science},
	author = {Wei, David and Rubio-Abadal, Antonio and Ye, Bingtian and Machado, Francisco and Kemp, Jack and Srakaew, Kritsana and Hollerith, Simon and Rui, Jun and Gopalakrishnan, Sarang and Yao, Norman Y. and Bloch, Immanuel and Zeiher, Johannes},
	month = may,
	year = {2022},
	pages = {716--720},
}

@article{joshi_observing_2022,
	title = {Observing emergent hydrodynamics in a long-range quantum magnet},
	volume = {376},
	issn = {0036-8075, 1095-9203},
	url = {https://www.science.org/doi/10.1126/science.abk2400},
	doi = {10.1126/science.abk2400},
	language = {},
	number = {6594},
	urldate = {2026-07-03},
	journal = {Science},
	author = {Joshi, M. K. and Kranzl, F. and Schuckert, A. and Lovas, I. and Maier, C. and Blatt, R. and Knap, M. and Roos, C. F.},
	month = may,
	year = {2022},
	pages = {720--724},
}

@article{tanaka_electron_1985,
	title = {Electron {Paramagnetic} {Resonance} in the {Quasi}-{One}-{Dimensional} {Jahn}-{Teller}-{Crystals}. {I}. {CsCuCl}$_{\textrm{3}}$},
	volume = {54},
	issn = {0031-9015, 1347-4073},
	url = {http://journals.jps.jp/doi/10.1143/JPSJ.54.4345},
	doi = {10.1143/JPSJ.54.4345},
	language = {},
	number = {11},
	urldate = {2026-07-03},
	journal = {Journal of the Physical Society of Japan},
	author = {Tanaka, Hidekazu and Iio, Katsunori and Nagata, Kazukiyo},
	month = nov,
	year = {1985},
	pages = {4345--4358},
}

@article{zhao2025superspin,
   title = {Superspin renormalization and slow relaxation in random spin systems},
   author = {Zhao, Yi J. and Garratt, Samuel J. and Moore, Joel E.},
   journal = {Phys. Rev. B},
   volume = {112},
   issue = {5},
   pages = {054436},
   numpages = {25},
   year = {2025},
   month = {Aug},
   publisher = {American Physical Society},
   doi = {10.1103/ppkd-dff1},
   url = {https://link.aps.org/doi/10.1103/ppkd-dff1}
}

@article{yao2014many,
  title={Many-body localization in dipolar systems},
  author={Yao, Norman Y and Laumann, Chris R and Gopalakrishnan, Sarang and Knap, Michael and Mueller, Markus and Demler, Eugene A and Lukin, Mikhail D},
  journal={Physical Review Letters},
  volume={113},
  number={24},
  pages={243002},
  year={2014},
  publisher={APS},
  url = {https://link.aps.org/doi/10.1103/PhysRevLett.113.243002}
}

@article{choi_robust_2020,
    title = {Robust {Dynamic} {Hamiltonian} {Engineering} of {Many}-{Body} {Spin} {Systems}},
    volume = {10},
    url = {https://link.aps.org/doi/10.1103/PhysRevX.10.031002},
    doi = {10.1103/PhysRevX.10.031002},
    number = {3},
    journal = {Physical Review X},
    publisher = {American Physical Society},
    author = {Choi, Joonhee and Zhou, Hengyun and Knowles, Helena S. and Landig, Renate and Choi, Soonwon and Lukin, Mikhail D.},
    month = jul,
    year = {2020},
    pages = {031002},
}

@article{mermin_absence_1966,
    title = {Absence of {Ferromagnetism} or {Antiferromagnetism} in {One}- or {Two}-{Dimensional} {Isotropic} {Heisenberg} {Models}},
    volume = {17},
    copyright = {http://link.aps.org/licenses/aps-default-license},
    issn = {0031-9007},
    url = {https://link.aps.org/doi/10.1103/PhysRevLett.17.1133},
    doi = {10.1103/PhysRevLett.17.1133},
    language = {},
    number = {22},
    urldate = {2026-07-31},
    journal = {Physical Review Letters},
    author = {Mermin, N. D. and Wagner, H.},
    month = nov,
    year = {1966},
    pages = {1133--1136},
}

@article{hohenberg_theory_1977,
    title = {Theory of dynamic critical phenomena},
    volume = {49},
    copyright = {http://link.aps.org/licenses/aps-default-license},
    issn = {0034-6861},
    url = {https://link.aps.org/doi/10.1103/RevModPhys.49.435},
    doi = {10.1103/RevModPhys.49.435},
    language = {},
    number = {3},
    urldate = {2024-06-16},
    journal = {Reviews of Modern Physics},
    author = {Hohenberg, P. C. and Halperin, B. I.},
    month = jul,
    year = {1977},
    pages = {435--479},
}

@book{sachdev_quantum_2023,
    edition = {1},
    title = {Quantum {Phases} of {Matter}},
    copyright = {https://www.cambridge.org/core/terms},
    isbn = {978-1-009-21271-7 978-1-009-21269-4},
    url = {https://www.cambridge.org/core/product/identifier/9781009212717/type/book},
    doi = {10.1017/9781009212717},
    urldate = {2026-07-31},
    publisher = {Cambridge University Press},
    author = {Sachdev, Subir},
    month = mar,
    year = {2023},
}

@book{giamarchi_quantum_2003,
    title = {Quantum {Physics} in {One} {Dimension}},
    isbn = {978-0-19-852500-4},
    url = {https://doi.org/10.1093/acprof:oso/9780198525004.001.0001},
    doi = {10.1093/acprof:oso/9780198525004.001.0001},
    publisher = {Oxford University Press},
    author = {Giamarchi, Thierry},
    month = dec,
    year = {2003},
}

@article{jepsen_spin_2020,
    title = {Spin transport in a tunable {Heisenberg} model realized with ultracold atoms},
    volume = {588},
    issn = {0028-0836, 1476-4687},
    url = {https://www.nature.com/articles/s41586-020-3033-y},
    doi = {10.1038/s41586-020-3033-y},
    language = {},
    number = {7838},
    urldate = {2026-07-31},
    journal = {Nature},
    author = {Jepsen, Paul Niklas and Amato-Grill, Jesse and Dimitrova, Ivana and Ho, Wen Wei and Demler, Eugene and Ketterle, Wolfgang},
    month = dec,
    year = {2020},
    pages = {403--407},
}

@article{hlubek_ballistic_2010,
    title = {Ballistic heat transport of quantum spin excitations as seen in {SrCuO}$_2$},
    volume = {81},
    copyright = {http://link.aps.org/licenses/aps-default-license},
    issn = {1098-0121, 1550-235X},
    url = {https://link.aps.org/doi/10.1103/PhysRevB.81.020405},
    doi = {10.1103/PhysRevB.81.020405},
    language = {},
    number = {2},
    urldate = {2026-07-31},
    journal = {Physical Review B},
    author = {Hlubek, N. and Ribeiro, P. and Saint-Martin, R. and Revcolevschi, A. and Roth, G. and Behr, G. and B\"{u}chner, B. and Hess, C.},
    month = jan,
    year = {2010},
    pages = {020405},
}

@article{braemer_cluster_2024,
    title = {Cluster truncated {Wigner} approximation for bond-disordered {Heisenberg} spin models},
    volume = {110},
    issn = {2469-9950, 2469-9969},
    url = {https://link.aps.org/doi/10.1103/PhysRevB.110.054204},
    doi = {10.1103/PhysRevB.110.054204},
    language = {},
    number = {5},
    urldate = {2026-07-31},
    journal = {Physical Review B},
    author = {Braemer, Adrian and Vahedi, Javad and G\"attner, Martin},
    month = aug,
    year = {2024},
    pages = {054204},
}

@article{schachenmayer_many-body_2015,
    title = {Many-{Body} {Quantum} {Spin} {Dynamics} with {Monte} {Carlo} {Trajectories} on a {Discrete} {Phase} {Space}},
    volume = {5},
    copyright = {http://creativecommons.org/licenses/by/3.0/},
    issn = {2160-3308},
    url = {https://link.aps.org/doi/10.1103/PhysRevX.5.011022},
    doi = {10.1103/PhysRevX.5.011022},
    language = {},
    number = {1},
    urldate = {2024-04-28},
    journal = {Physical Review X},
    author = {Schachenmayer, J. and Pikovski, A. and Rey, A. M.},
    month = feb,
    year = {2015},
    pages = {011022},
}

@article{zhang_reporter-spin-assisted_2023,
    title = {Reporter-{Spin}-{Assisted} {$T_1$} {Relaxometry}},
    volume = {19},
    issn = {2331-7019},
    url = {https://link.aps.org/doi/10.1103/PhysRevApplied.19.L031004},
    doi = {10.1103/PhysRevApplied.19.L031004},
    language = {},
    number = {3},
    urldate = {2026-09-01},
    journal = {Physical Review Applied},
    author = {Zhang, Zhiran and Joos, Maxime and Bluvstein, Dolev and Lyu, Yuanqi and Bleszynski Jayich, Ania C.},
    month = mar,
    year = {2023},
    pages = {L031004},
}

@article{laraoui_high-resolution_2013,
    title = {High-resolution correlation spectroscopy of {$^{13}$C} spins near a nitrogen-vacancy centre in diamond},
    volume = {4},
    copyright = {2013 Springer Nature Limited},
    issn = {2041-1723},
    url = {https://www.nature.com/articles/ncomms2685},
    doi = {10.1038/ncomms2685},
    language = {},
    number = {1},
    urldate = {2024-03-11},
    journal = {Nature Communications},
    publisher = {Nature Publishing Group},
    author = {Laraoui, Abdelghani and Dolde, Florian and Burk, Christian and Reinhard, Friedemann and Wrachtrup, Jörg and Meriles, Carlos A.},
    month = apr,
    year = {2013},
    pages = {1651},
}

@article{liu_surface_2022,
    title = {Surface {NMR} using quantum sensors in diamond},
    volume = {119},
    issn = {0027-8424, 1091-6490},
    url = {https://pnas.org/doi/full/10.1073/pnas.2111607119},
    doi = {10.1073/pnas.2111607119},
    language = {},
    number = {5},
    urldate = {2026-09-01},
    journal = {Proceedings of the National Academy of Sciences},
    author = {Liu, Kristina S. and Henning, Alex and Heindl, Markus W. and Allert, Robin D. and Bartl, Johannes D. and Sharp, Ian D. and Rizzato, Roberto and Bucher, Dominik B.},
    month = feb,
    year = {2022},
    pages = {e2111607119},
}

@article{schuckert_nonlocal_2020,
    title = {Nonlocal emergent hydrodynamics in a long-range quantum spin system},
    volume = {101},
    issn = {2469-9950, 2469-9969},
    url = {https://link.aps.org/doi/10.1103/PhysRevB.101.020416},
    doi = {10.1103/PhysRevB.101.020416},
    language = {},
    number = {2},
    urldate = {2026-09-04},
    journal = {Physical Review B},
    author = {Schuckert, Alexander and Lovas, Izabella and Knap, Michael},
    month = jan,
    year = {2020},
    pages = {020416},
}

@article{kloss_spin_2019,
    title = {Spin transport in a long-range-interacting spin chain},
    volume = {99},
    issn = {2469-9926, 2469-9934},
    url = {https://link.aps.org/doi/10.1103/PhysRevA.99.032114},
    doi = {10.1103/PhysRevA.99.032114},
    language = {},
    number = {3},
    urldate = {2026-09-04},
    journal = {Physical Review A},
    author = {Kloss, Benedikt and Bar Lev, Yevgeny},
    month = mar,
    year = {2019},
    pages = {032114},
}

@article{nandkishore_lifetimes_2021,
    title = {Lifetimes of local excitations in disordered dipolar quantum systems},
    volume = {103},
    issn = {2469-9950, 2469-9969},
    url = {https://link.aps.org/doi/10.1103/PhysRevB.103.134423},
    doi = {10.1103/PhysRevB.103.134423},
    language = {},
    number = {13},
    urldate = {2024-05-07},
    journal = {Physical Review B},
    author = {Nandkishore, Rahul and Gopalakrishnan, Sarang},
    month = apr,
    year = {2021},
    pages = {134423},
}

@article{stasiuk_disorder-induced_2025,
    title = {Disorder-{Induced} {Anomalous} {Diffusion} in a {3D} {Spin} {Network}},
    copyright = {Creative Commons Attribution 4.0 International},
    url = {https://arxiv.org/abs/2510.09549},
    author = {Stasiuk, Andrew and Heller, Garrett and Berkey, Lance and Xing, Bo and Cappellaro, Paola},
    year = {2025},
    journal = {arXiv preprint},
}

@article{selco_continuous-time_2026,
    title = {Continuous-{Time} {Random} {Walk} {Description} of {Anomalous} {Spin} {Transport} in {Dilute} {Dipolar} {Networks}},
    copyright = {Creative Commons Attribution 4.0 International},
    url = {https://arxiv.org/abs/2607.22626},
    author = {Selco, Cooper M. and Bengs, Christian and Ajoy, Ashok},
    year = {2026},
    journal = {arXiv preprint},
}

@article{postelnicu_harnessingdiamond_2026,
    title = {{Harnessing} {Diamond} {Surface} {Features} for {Dense} {andAligned} {NV} {Ensembles}},
    volume = {26},
    copyright = {https://doi.org/10.15223/policy-029},
    issn = {1530-6984, 1530-6992},
    url = {https://pubs.acs.org/nalefd/article/26/34/11314/5267258/Harnessing-Diamond-Surface-Features-for-Dense-and},
    doi = {10.1021/acs.nanolett.6c02155},
    language = {},
    number = {34},
    urldate = {2026-09-05},
    journal = {Nano Letters},
    author = {Postelnicu, Eveline and Wyatt, Lillian B. Hughes and Nguyen, Tri and Meynell, Simon A. and Jilly, Christine and Wallace, Paul and Barnum, Andrew and Bleszynski Jayich, Ania and Mukherjee, Kunal},
    month = sep,
    year = {2026},
    pages = {11314--11323},
}

\onecolumngrid
\begin{center}\textbf{End Matter}\end{center}
\twocolumngrid

\section*{Ensemble averaging}
We begin by reiterating that the correlation spectroscopy signal consists of a configurational average over all NV-P1 positions and an average over infinite-temperature initial spin states.
In all measurements, we assume that the experimental signal we measure \emph{converges fully} under both forms of averaging.
The choice of infinite-temperature density matrix is justified by $k_B T \gg \gamma_e B$ in our experiments, where $\gamma_e = \SI{2.8}{\mega\hertz\per G}$ is the electron gyromagnetic ratio, $B\sim\SI{350}{G}$ is our magnetic field, and $T = \SI{296}{\kelvin}$ is the temperature of our sample.

The assumption that our configurational averaging is complete holds for the following two reasons.
First, our confocal spot measures an ensemble of NV centers, each of which sees a distinct local P1 configuration.
Second, the $\pi$-pulses in the correlation measurement only recouple one P1 spectral group. 
At room temperature, each P1 center hops into and out of the targeted spectral group over the course of the measurement with a characteristic timescale of $\sim\SI{1}{\second}$~\cite{xiao_proposal_2016, grinolds_subnanometre_2014, ammerlaan_reorientation_1981}.
Taken together, although each individual NV in the ensemble may sense a positionally static P1 ensemble during a single measurement shot, ensemble NV measurement and P1 hopping ensure that a single NV averages over $\sim 10^6$ distinct local P1 positional configurations over the entire experiment. 
We note that varying charge-state initialization configurations~\cite{wang_manipulating_2023, monge_beyond_2025} across shots may contribute to averaging as well.

\section*{Spin survival probability approximation to correlation spectroscopy signal}

In this section, we derive and elaborate on the claim that the correlation spectroscopy signal approximates the spin survival probability $\overline{\langle B_\text{P1}(t) B_\text{P1}(0) \rangle} \approx \langle p_i^z(t) p_i^z(0) \rangle \propto S_p(t)$ in the main text.
We reproduce Eq.~(\ref{eq:autocorrelation}) in the main text here:
\begin{equation}\label{eq:autocorrelation_rep}
    \overline{\langle B_\text{P1}(t) B_\text{P1}(0) \rangle} = \overline{\left\langle \sum_{i} \frac{J_0^2}{r_i^6} p^z_i(t) p^z_i(0) \right\rangle} + \overline{\left\langle \sum_{i \neq j} \frac{J_0^2}{r_i^3 r_j^3} p^z_i(t) p^z_j(0) \right\rangle}.
\end{equation}
The configurational average allows us to analytically bound the relative sizes of the two expectation values in Eq.~(\ref{eq:autocorrelation_rep}) over a random distribution of NV-P1 distances $r_i$ and spin states $p_i^z(t)$.

Consider the first term. 
We assume that the NVs are uniformly distributed, dilute, off-resonant (i.e., Ising interaction only) probes of P1 dynamics. 
As such, the dynamics in the $z$ polarization of P1 spin $i$ should be uncorrelated with its proximity $r_i$ to the probe NV. 
We may therefore separate the averages:
\begin{align}
    \overline{\left\langle \sum_{i} \frac{J_0^2}{r_i^6}\, p^z_i(t)\, p^z_i(0) \right\rangle}
    &= \sum_{i} \overline{\frac{J_0^2}{r_i^6}}\;\; \overline{\left\langle p^z_i(t)\, p^z_i(0) \right\rangle} \\
    &= N \overline{J^2} \ \left\langle p^z_i(t)\, p^z_i(0) \right\rangle,
\end{align}
where we have also invoked linearity and the fact that each spin $i$ is identically distributed. 
We define the mean-squared interaction strength $\overline{J^2} = \overline{J_0^2/r_i^6}$.

Consider the second term. 
While the product of two spin distances $r_i$ and $r_j$ is not obviously independent of the cross-correlations $p_i^z(t)p_j^z(0)$ that build up between them, we may still bound the norm of this term (via the triangle inequality, and then by noting that $-1 \leq \overline{\langle p^z_i(t)p^z_j(0) \rangle} \leq 1$):
\begin{align}
    \left|\overline{\left\langle \sum_{i \neq j} \frac{J_0^2}{r_i^3 r_j^3}\, p^z_i(t)\, p^z_j(0) \right\rangle}\right|  
    &\leq \sum_{i\neq j}\overline{ \frac{J_0^2}{r_i^3  r_j^3} \left\langle|p_i^z(t)p_j^z(0)| \right\rangle} \nonumber \\ 
    & \leq \sum_{i\neq j} \overline{ \frac{J_0^2}{r_i^3 r_j^3}} \nonumber \\
    & = \sum_{i\neq j} \overline{\frac{J_0}{r_i^3}}\ \overline{\frac{J_0}{r_j^3}} \nonumber\\
    & = N (N-1) \overline{J}^2,
\end{align}
where we have defined a mean interaction strength $\overline{J} = \overline{J_0/r_i^3}$ and again exploited the positional independence of spins $i$ and $j$.

We can now analytically bound the size of the cross-correlation term relative to the autocorrelation term by calculating the ratio of $N\overline{J}^2$ to $\overline{J^2}$. 
The constituent integrals are evaluated in Supplemental Material~\cite{sup_note} and provide tight bounds that justify the correlation measurement approximation even for small $N$. 

For additional theoretical discussion, here we shift to the large-$N$ limit (holding the 1D spin density $\rho = N/L$ constant). 
In this limit, the ratio we wish to evaluate reduces to
\begin{equation}\label{eq:IPR_intro}
    \frac{N\overline{J}^2}{\overline{J^2}} \sim \rho r_\text{min},
\end{equation} 
which can be interpreted as follows. 

The quantity $r_\text{min}$ can be thought of as the underlying lattice constant of a partially filled ordered lattice, often characterized by a filling factor $f$. 
Under this interpretation, $\rho r_\text{min} = N/N_\text{max}\sim f$, where $N_\text{max} = L/r_\text{min}$ is the maximal number of spins placed in volume $L$ subject to the minimum spacing constraint. 
In this light, our approximation can be thought of as a dilute lattice (i.e., $f\ll1$) approximation.
We note the quantity $N\overline{J}^2/ \overline{J^2}$ has also appeared in literature as ``effective coordination number'' that describes the fraction of P1 centers that contribute the most signal to the sensor NV~\cite{gao_signal_2025}.
In other contexts, this quantity also controls the build-up of pairwise correlations in dipolar interacting systems~\cite{lacelle_dipolar_1995, mukherjee_influence_2024, hazzard_many-body_2014}. 

In summary, we recover the approximation quoted in the main text,
\begin{equation}
    \frac{\enspace\overline{\langle B_\text{P1}(t) B_\text{P1}(0) \rangle}\enspace}{\overline{\langle B_\text{P1}(0) B_\text{P1}(0) \rangle}} \approx S_p(t),
\end{equation}
where $S_p(t) \equiv \overline{\langle p^z_i(t)\, p^z_i(0) \rangle}$ is the single-spin autocorrelation function.

\begin{figure}
\includegraphics{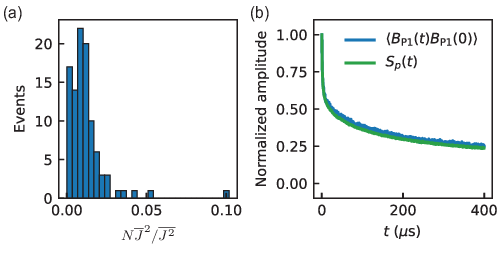}
\caption{(a) Histogram of simulated $N\overline{J}^2/\overline{J^2}$ from 100 random P1 positional configurations, with a sensor NV located at the origin. 
For each configuration, 100 P1 centers are generated over a region defined by the measured P1 density. 
(b) Comparison between $\langle B_\text{P1}(t) B_\text{P1}(0) \rangle$ and $S_p(t)$ of P1 dynamics under clustered dTWA simulations. 
Near-agreement between $\langle B_\text{P1}(t) B_\text{P1}(0) \rangle$ and $S_p(t)$ is observed, confirming our analytical derivation.
However, we note that numerical simulations do not capture the experimentally measured spin dynamics.}
\label{fig:cross_terms}
\end{figure}

Using numerical methods, we further benchmark the survival probability approximation for a realistically chosen filling factor.
First, we numerically calculate $N/N_{\text{max}}$ in a finite volume by randomly scattering P1s over a fixed geometry and computing the distribution of dipolar couplings to a central NV.
Given the density of the P1 ensemble~\cite{sup_note}, we find $N/N_{\text{max}} \approx 2\%$ [Fig.~\ref{fig:cross_terms}(a)], consistent with analytical estimates.
Next, we perform semiclassical quantum dynamical simulations using the clustered discrete truncated Wigner approximation (dTWA) over $10^4$ P1 configurations~\cite{schachenmayer_many-body_2015,braemer_cluster_2024}, finding close numerical agreement between simulated $\overline{\langle B_\text{P1}(t) B_\text{P1}(0) \rangle}$ and simulated $S_p(t)$ [Fig.~\ref{fig:cross_terms}(b)], as expected.

Finally, we note that $J_0$ is held constant throughout the derivation, as is the case for a 1D ensemble where the angular contribution is constant.
In 2D (100) and 3D, by contrast, averaging over the angular dependence of $J$ significantly simplifies the derivation, resulting in $\overline{J} = 0$.
Therefore, the cross term in Eq.~(\ref{eq:autocorrelation}) is suppressed independently of any filling factor arguments, and correlation spectroscopy as a probe of spin survival probability extends naturally to higher-dimensional geometries.

\begin{figure}
    \includegraphics{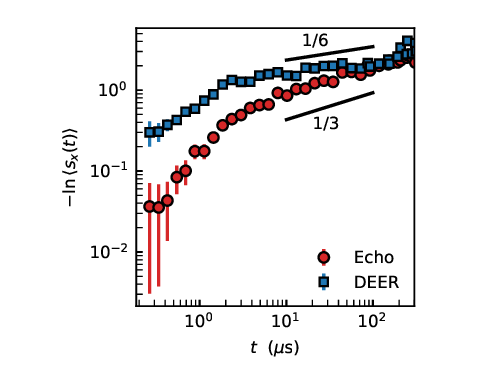}
    \caption{
    Undriven echo and DEER measurements.
    A late-time $1/6$ stretched exponential portion is observed for more than two decades of time past $t\sim \SI{3}{\micro\second}$ in DEER, but not in the echo.
    This is consistent with P1s undergoing different dynamics in the two measurements.
    }
    \label{fig:undriven}
\end{figure}

\section*{Remarks on echo and DEER measurements in the absence of stochastic driving}
One main difference in dimensionality estimation in this work compared to prior works~\cite{dwyer_probing_2022,hughes_two-dimensional_2023, davis_probing_2023} is the utilization of stochastic driving on the dark P1 spins.
Conventionally, comparisons between echo and DEER measurements are performed by assuming that the underlying dynamics of the dark spins, characterized by the single-spin autocorrelation functions, are the same in both measurements.
In such a situation, echo and DEER measurements can be compared by analytically accounting for the differences in their respective filter functions~\cite{dwyer_probing_2022, hughes_two-dimensional_2023,yu_engineering_2026}.

However, the assumption that the spin dynamics are insensitive to the pulse sequence is not true in the dipolar interacting limit.
As demonstrated in Fig.~\ref{fig:correlation}, the P1 dynamics are noticeably modified under dynamical decoupling that removes on-site disorder.
Similarly, in a DEER measurement without stochastic driving (``undriven DEER''), the addition of a $\pi$-pulse on the P1 ensemble also decouples on-site static disorder, accelerating depolarization and resulting in undriven DEER and undriven echo measurements that capture different spin dynamics.
In Fig.~\ref{fig:undriven}, we show the undriven echo and DEER measurements on our 1D interacting spot. 
A pronounced late-time regime is observed in the DEER decay, but not in the echo, pointing to the fact that the limiting spin bath dynamics are quite different in the two cases.

Stochastic driving, in this case, circumvents this issue by imposing dynamics that are unchanged between the two measurements, providing a more fiducial method of comparison.

~\nocite{eichhorn_optimizing_2019}

\end{document}